\documentclass[a4paper,12pt, preprint, numbers,sort&compress]{elsarticle}
\usepackage{textcomp}
\usepackage{graphicx}
\usepackage[usenames]{color}
\usepackage{natbib}
\usepackage{amssymb}

\begin{document}
%\begin{frontmatter}

\title{Organic charge-transfer phase formation in thin films of the BEDT-TTF/TCNQ donor-acceptor system}
\author[uni]{Vita Levitan\corref{cor1}}
\ead{levitan@physik.uni-frankfurt.de}
\author[uni]{Kerstin Keller}
\author[uni]{Michael Huth}
\cortext[cor1]{Corresponding author}
\address[uni]{Physikalisches Institut, Goethe-Universit\"{a}t,\\Max-von-Laue-Strasse 1, 60438 Frankfurt am Main, Germany}

\begin{abstract}
We have performed charge transfer phase formation studies on the donor/ac\-ceptor system bis(ethylendithio)tetrathiafulvalene\,(BEDT-TTF)/tetracyano\-quinodimethane\,(TCNQ) by means of physical vapor deposition. We prepared donor/acceptor bilayer structures on glass and Si(100)/SiO$_2$ substrates held at room temperature and analyzed the layer structures by optical microscopy, X-ray diffraction and focused ion beam cross sectioning before and after annealing. We found clear evidence for the formation of a charge transfer phase during the annealing procedure. For the as-grown samples we could not detect the occurrence of a charge transfer phase. X-ray diffraction indicated that the monoclinic variant of the (BEDT-TTF)-TCNQ was formed. This was further corroborated by single-source evaporation experiments from pre-reacted (BEDT-TTF)-TCNQ obtained from solution growth, and in particular from co-evaporation experiments of (BEDT-TTF)-TCNQ and TCNQ. In the course of these experiments we found that $(0\ell\ell)$-oriented BEDT-TTF layers can be prepared on $\alpha$-Al$_2$O$_3$ $(11\bar{2}0)$ substrates at about 100~$^{\circ}$C using (BEDT-TTF)-TCNQ as source material. We speculate that due to its high vapor pressure the TCNQ component serves as a carrier gas for BEDT-TTF vapor phase transport.
\end{abstract}

\begin{keyword}
Organic thin films \sep charge transfer \sep BEDT-TTF \sep TCNQ
\PACS 68.37.-d \sep 68.55.am \sep 81.15.Kk
\end{keyword}
%\end{frontmatter}
\maketitle

\section{Introduction}
Organic materials attract significant attention since many years because of their interesting physical properties rendering them promising for applications \cite{Inokuchi_rev,Saito_rev,Kawamoto} and for basic research on electronic correlation phenomena \cite{Toyota07}. The interest in electronic properties of organic materials arose shortly after the first paper by Inokuchi \textit{et al.} \cite{Inokuchi_nature} who discovered unusually high electrical conductivity in the perylene-bromine complex. Before this discovery the organic materials were considered to be insulators. Within the class of organic materials charge transfer (CT) compounds are attractive since they hold the promise to significantly extend the presently developing organic electronics. They furthermore represent model systems for studying the interplay of electronic correlation effects, frustration and dimensionality \cite{Toyota07}.

Within the many different organic charge transfer compounds which have been studied during the last decades, the complex (BEDT-TTF)-TCNQ is particularly interesting \cite{Strange_ET_TCNQ, Metal-Insulator_ET-TCNQ, FET_casting}. It exists in three different structural variants, namely as monoclinic semiconductor \cite{monoclinic}, triclinic semiconductor $\beta$\textasciiacute $ $ \cite{Semi_1986} or triclinic metal $\beta$\textacutedbl$ $ \cite{Metallic_beta2, Mori_beta_beta2}. The triclinic semiconductor phase experiences a correlation driven metal-insulator transition at 330~K \cite{Semi_1986, Metal-Insulator_ET-TCNQ}, while the triclinic metallic $\beta$\textacutedbl $ $ phase shows anomalies in its electrical conductivity, which are assumed to be of magnetic origin \cite{Strange_ET_TCNQ}.

Presently the preparation of organic compounds in form of thin films is a necessary step in the construction of organic material based devices, such as organic solar cells, electrochemical cells, organic light emitting diodes and organic field-effect transistors (for review see \cite{Koch_devices, Organic_molecular, Forrest_review, Organic_single_transistors} and references therein). A common method for the formation of organic compounds is the solution growth method, which allows to obtain single crystalline substances (see e.g. Refs.~\cite{monoclinic, Semi_1986}). An adopted variant of the solution growth for preparing thin films is the drop-casting method \cite{FET_casting, Ambipolar_FET}. As the solvent evaporates in the course of the crystallization process, solvent molecules may get incorporated in the forming material, influencing its structural and electronic properties. This must be considered as a disadvantage of this technique. Inherent to the solvent-based methods is a second disadvantage for poorly soluble donor or acceptor species. In this case the nucleation rate is high which results in a disordered agglomerate of small crystals.

For the formation of ordered thin films of CT systems by vapor deposition of the donor and acceptor species the above mentioned disadvantages do not occur, but other problems arise. {\it A priori} it is not clear whether a CT complex will form without the presence of a solvent. In solvent-based growth the dielectric properties of the solvent reduce the activation barrier for charge transfer. Moreover, careful selection of the solvent allows for biasing the nucleation rate towards the desired structural variant of the growing CT complex. In vapor-phase growth of thin films the substrate's dielectric properties may play a significant role in CT, but this role is yet not well studied. As a starting point for studying CT phase formation, the bilayer sequential growth of acceptor/donor thin layers followed by a careful analysis of the acceptor/donor interface region can be efficiently used.

In spite of the interesting properties of (BEDT-TTF)-TCNQ, a systematic study on its phase formation via the vapor phase route is still missing. In the present work we have used physical vapor deposition to form acceptor (TCNQ)-donor(BEDT-TTF) bilayer films followed by post growth annealing. We studied the phase formation of (BEDT-TTF)-TCNQ at the layers' interface by inspecting the interface region employing focused ion beam (FIB) cross sectioning. We furthermore studied the formation of (BEDT-TTF)-TCNQ by co-evaporating BEDT-TTF and TCNQ, as well as using (BEDT-TTF)-TCNQ crystallites from solution growth as source material. We found that a CT phase does form at the interface region of bilayer samples under proper annealing conditions. We also obtained CT phase layers via the co-evaporation route without annealing. The composition of the CT crystals was analyzed by energy dispersive X-ray spectroscopy (EDX). Employing X-ray diffraction we identified the CT phase as the monoclinic variant.

\section{Experimental Methods}
BEDT-TTF/TCNQ bilayers of various thicknesses were prepared by sequential physical vapor deposition of a TCNQ layer and a BEDT-TTF layer at a background pressure of $3\times~10^{-7}$~mbar or less. The materials were sublimated from low-temperature effusion cells using quartz crucibles. TCNQ was deposited at 110~$^{\circ}$C and BEDT-TTF at 155~$^{\circ}$C. The growth rate for both layers was about 2\,nm/s. The cell temperatures were measured by Ni-NiCr-thermocouples, coupled to the heated body of the effusion cells by copper wool. We could realize these rather high growth rates by having a small distance between the effusion cell lid and substrate of about 10~mm. The TCNQ layer was deposited first because it has a higher vapor pressure than BEDT-TTF at any given temperature and tends to desorb from a substrate at even moderately elevated temperatures. As substrate materials for the bilayers we used glass microscope slides and Si(100)/SiO$_{2}$ (300nm) substrates with and without Au template layers. During growth the substrates were held at room temperature. The post-growth annealing of the samples was done in $^{4}$He inert gas atmosphere and in vacuum at temperatures varying from 60~$^{\circ}$C to 80~$^{\circ}$C for up to 48\,hrs employing a Peltier element on which the substrates were mounted.

In a second series of experiments, performed in a different preparation chamber, the method of co-evaporation was employed. We used BEDT-TTF and TCNQ, as well as pre-formed (BEDT-TTF)-TCNQ crystals and TCNQ as co-evaporants. The (BEDT-TTF)-TCNQ crystals (typically a mixture of the monoclinic and $\beta$'-phases) used as source material were obtained from solution growth with THF or Dichlormethane as solvent as detailed in the literature \cite{monoclinic, Semi_1986}. For these experiments $\alpha$-Al$_{2}$O$_{3}$ in $(11\bar{2}0)$-orientation (a-plane) and Si(100)/SiO$_{2}$ (300nm) were used. Thin films on the two substrates were prepared simultaneously under identical conditions. The sublimation temperatures were 130-138~$^{\circ}$C for TCNQ and 136~$^{\circ}$C for (BEDT-TTF)-TCNQ. The substrate temperature was controlled by a resistive heater and was varied between room temperature and 100~$^{\circ}$C.

A Leica DM 4000M optical microscope was used for optical inspection. X-ray diffractometry was done employing a Bruker D8 diffractometer with a Cu anode in parallel beam mode using a Goebel mirror. Scanning electron microscopy (SEM) was done with an FEI xT Nova NanoLab 600 microscope at 5\,kV and a beam current of 98\,pA. A Ga ion source FIB operating at 30\,kV with beam currents between 30\,pA and 1\,nA was used for the preparation of cross sections. We employed energy-dispersive X-ray spectroscopy at 5~keV for determining the composition of the grown structures.
\section{Results}
\subsection{Characterization of bilayers} \label{ssec:char}
\begin{figure}[t]
%{0.5\linewidth}
\includegraphics[width=1\textwidth]{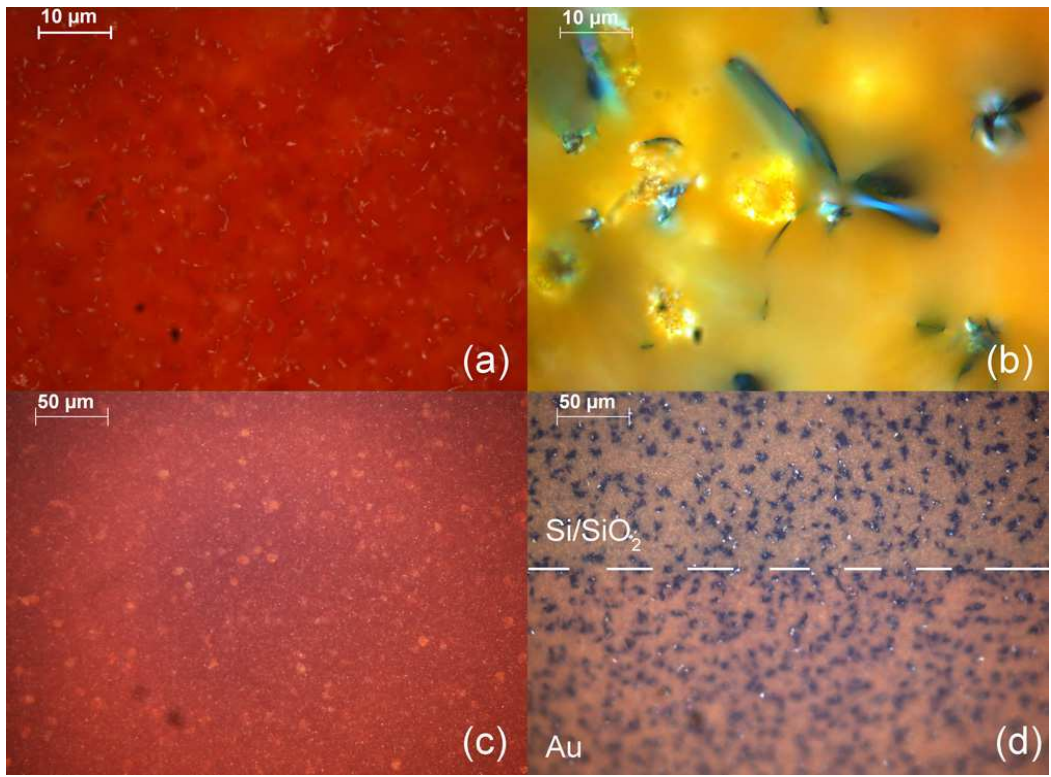}
\caption{Optical microscopy images of (BEDT-TTF)-TCNQ bilayer samples. Top row: bilayer grown on a glass substrate. Bottom row: bilayer grown on a Si/SiO$_{2}$ substrate one half of which is covered with a gold layer. Images (a) and (c) show the bilayer after the deposition of the BEDT-TTF layer on top of the TCNQ layer. Images (b) and (d) show the bilayer after the post growth annealing. The dashed line indicates the border between the bare Si/SiO$_{2}$ and Au-covered substrate surface. Images (a) and (b) were taken at 1000$\times$ magnification, while images (c) and (d) were taken at 200$\times$ magnification.}
\label{fig:optical}
\end{figure}

The bilayer samples were inspected by optical microscopy at different stages of the experiment, i.~e.\ after the deposition of each layer and after the post growth annealing. Typical images, as recorded after the deposition of the second layer and after the post growth annealing of a bilayer grown on a glass substrate, are shown in Figs.~\ref{fig:optical}a and \ref{fig:optical}b, respectively. Figure \ref{fig:optical}b shows that after the post growth annealing black crystals are formed originating from the interior of the top BEDT-TTF-layer. This is a first indication for a phase transformation in the system. The same observation was made for bilayers grown on Si(100)/SiO$_{2}$ (300nm) substrates one half of which had previously been covered by a gold layer, as is again evident from optical microscopy performed on the as-grown bilayer (see Fig.~\ref{fig:optical}c) and for the annealed bilayer (see Fig.~\ref{fig:optical}d).

\begin{figure}[t]
\includegraphics[width=1\textwidth]{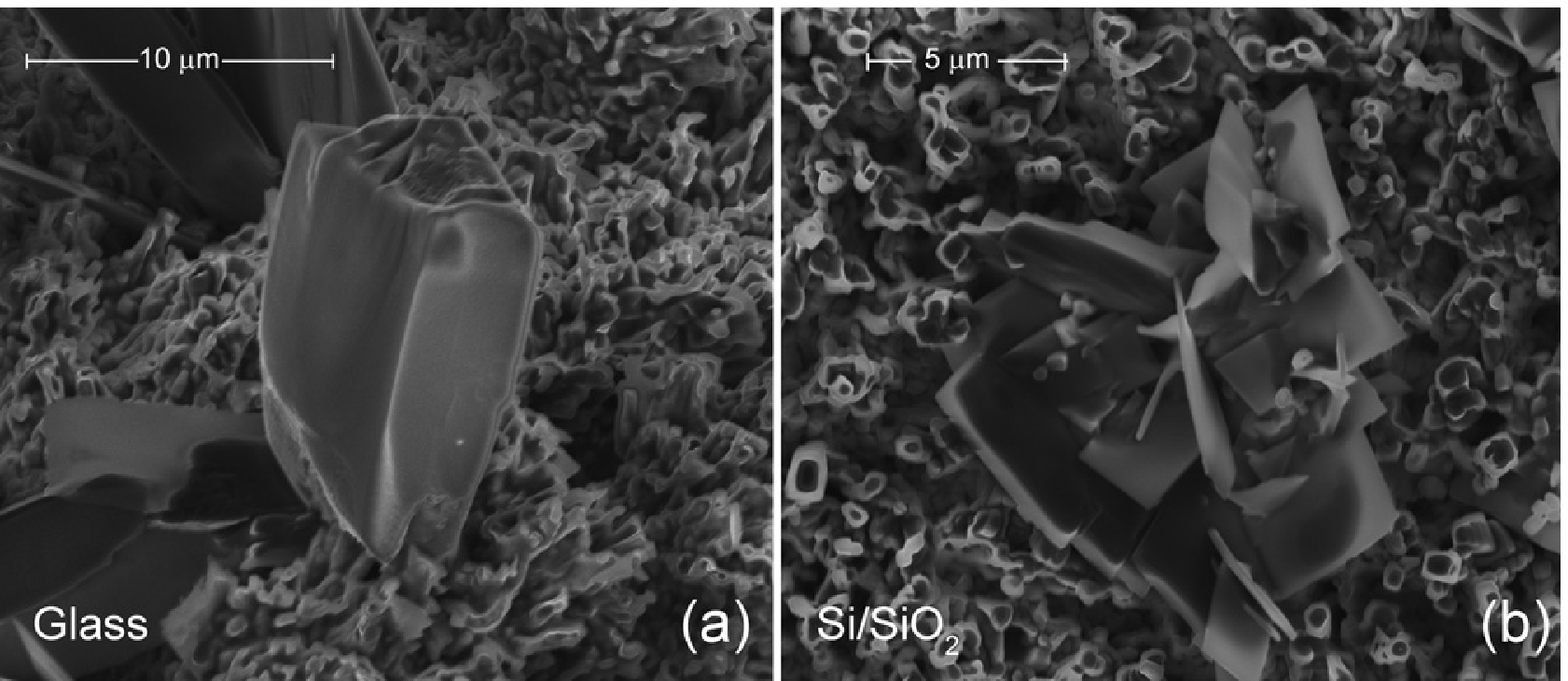}
\caption{(a) SEM image of individual (BEDT-TTF)-TCNQ crystal formed in the course of the post-growth annealing procedure of a bilayer grown on a glass substrate. The characteristic size of the crystal is $\sim$10~$\mu$m, as indicated in the figure. (b) A group of (BEDT-TTF)-TCNQ crystals which had been formed during the post-growth annealing of a bilayer grown on a Si(100)/SiO$_{2}$ substrate. The characteristic size of one crystal in this case is $<$5~$\mu$m, as indicated in the figure.}
\label{fig:electron2}
\end{figure}

For bilayers grown on the different substrate materials the size and the morphology of the crystallites formed after annealing were investigated by SEM. The characteristic size of crystals grown from layers on the glass substrate is $\gtrsim $10$\mu$m, as is exemplarily shown in Fig.~\ref{fig:electron2}a. The crystals have typically a prismatic shape and are oriented inclined to the film surface, as also follows from optical microscopy (see Fig.~\ref{fig:optical}b). Crystals obtained on Si(100)/SiO$_2$ and Si(100)/SiO$_2$/Au substrates show a plate-like shape and their characteristic size is smaller than 5~$\mu$m (see Fig.~\ref{fig:electron2}b). Note that for the Au-covered and bare part of the Si/SiO$_2$ substrate the crystals' average size and inclination to the surface do not appear to depend systematically on the underlying substrate surface. We speculate that the observed morphological differences between the crystallites formed on glass and Si/SiO$_2$ are due to the different rms-roughnesses of the respective substrate surfaces. From atomic force microscopy in non-contact mode we determined the rms-roughnesses to be 6~nm and 0.6~nm (scan range $10\times 10$~$\mu$m$^2$) for the glass and Si/SiO$_2$ substrates, respectively.

\begin{figure}[htb]
\includegraphics[width=1\textwidth]{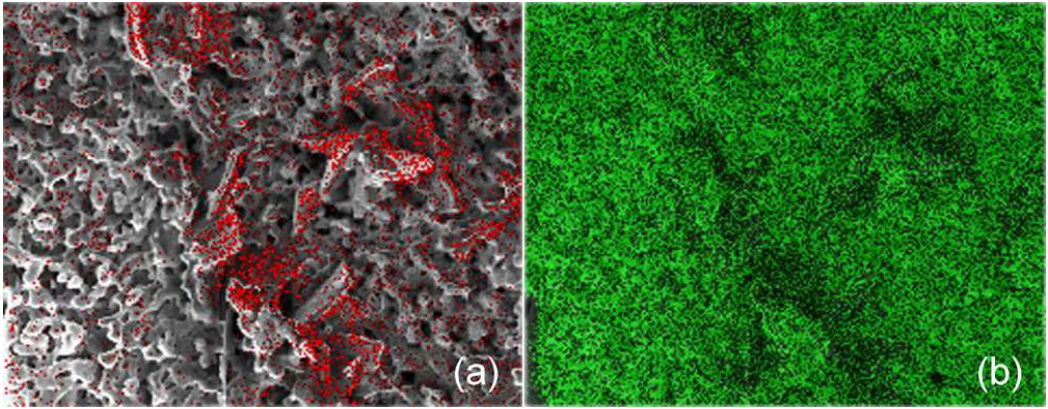}
\caption{EDX element maps. (a) Overlay of the nitrogen element map (K-line) with a SEM image of a crystallite group protruding from the annealed bilayer grown on Si/SiO$_2$. Nitrogen indicates the presence of TCNQ molecules. (b) Overlay of the sulphur element map (K-line) with the SEM image of the same crystallite group. Sulphur indicates the presence of BEDT-TTF molecules.}
\label{fig:EDX_map}
\end{figure}

In the next step we aim for identifying the chemical composition of the black crystallites formed after annealing. We used EDX to determine the abundance of sulphur, nitrogen and carbon of the crystallites in the bilayer grown on the Si/SiO$_2$ substrate as S:N:C=21.0:9.8:69 (in atomic percents) with an estimated error margin of $\pm 0.5$ for each component. The abundance of hydrogen atoms cannot be quantified by this method. The expected values for (BEDT-TTF)-TCNQ (C$_{22}$~H$_{12}$~N$_{4}$~S$_{8}$) are 23.5:11.8:64.7. The measured proportion of carbon is higher than expected, which we attribute to carbon deposition induced by the electron beam which dissociates adsorbed residual gases in the course of the EDX measurement. The ratio of sulphur to nitrogen in this measurement is 2.15$\pm$~0.15 which is, within the estimated error margin, in accordance with the expected ratio of 2.0 for the (BEDT-TTF)-TCNQ compound. We performed supplementary EDX element mapping experiments to identify regions of high abundance of nitrogen and sulphur. As follows from the EDX element maps overlaid with the SEM micrograph of a group of crystallites formed after annealing (see Fig.\ref{fig:EDX_map}), the enhanced abundance of nitrogen atoms coincides with the location of the crystallite group. The sulphur element map shows a rather homogenous distribution over the complete bilayer surface. From these observations we conclude that TCNQ molecules diffuse through the BEDT-TTF top layer during annealing. The element maps furthermore support the assumption that the black crystallites represent a CT phase of (BEDT-TTF)-TCNQ.

\begin{figure}[t]
%{0.5\linewidth}
\includegraphics[width=1\textwidth]{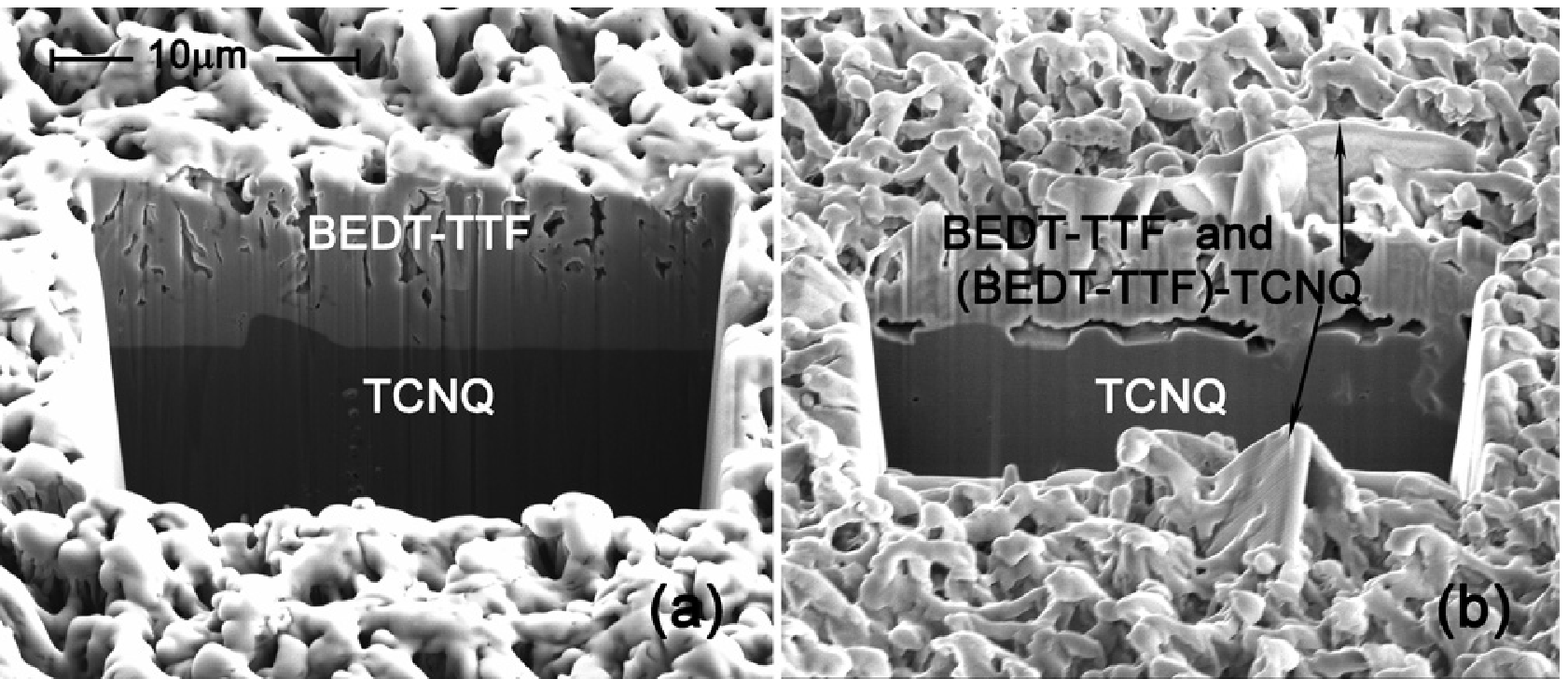}
\caption{SEM images of FIB cross section through the bilayers grown on a Si(100)/SiO$_2$ substrate, (a) before and (b) after the post-growth annealing. The images were recorded for the sample being inclined to the electron beam by 52$^{\circ}$. BEDT-TTF/TCNQ layers and (BEDT-TTF)-TCNQ crystals are indicated. The scale bar applies to both panels.}
\label{fig:fib}
\end{figure}

In order to obtain additional information about the TCNQ diffusion and CT phase formation process we prepared
FIB cross sections of the as-grown bilayer and for the same bilayer after post-growth annealing (see Figs.~\ref{fig:fib}a-b). In the cross sections the individual TCNQ and BEDT-TTF layers can be easily identified, with the BEDT-TTF layer appearing brighter because the average atomic number for this layer is higher than for the TCNQ layer. It is also apparent that the TCNQ layer is rather smooth whereas the BEDT-TTF layer is rough. In the as-grown sample the interface between the TCNQ and BEDT-TTF layers is smooth and planar. After annealing cavities have been formed at the layers' interface (see Fig.~\ref{fig:fib}b), which we attribute to diffusive transport of the highly mobile TCNQ molecules into the rough, capillaric BEDT-TTF top layer, in the course of which the CT phase formation occurred. The thus formed CT crystals grow upward from the interface region and extend beyond the BEDT-TTF top layer.

From the cross sections we determined the thickness of the top BEDT-TTF layer before and after annealing. We note a reduction of the thickness from about 6.4~$\mu$m to 3.3~$\mu$m. For TCNQ we can only specify a lower limit of 6.5~$\mu$m for the layer thickness, because we did not succeed to extend the cross section depth into the substrate surface. We attribute the reduction in the BEDT-TTF layer thickness to the CT phase formation and also to partial desorption of BEDT-TTF from the rough top-layer surface during the annealing process.

\begin{figure}[t]
%{0.5\linewidth}
\includegraphics[width=1\textwidth]{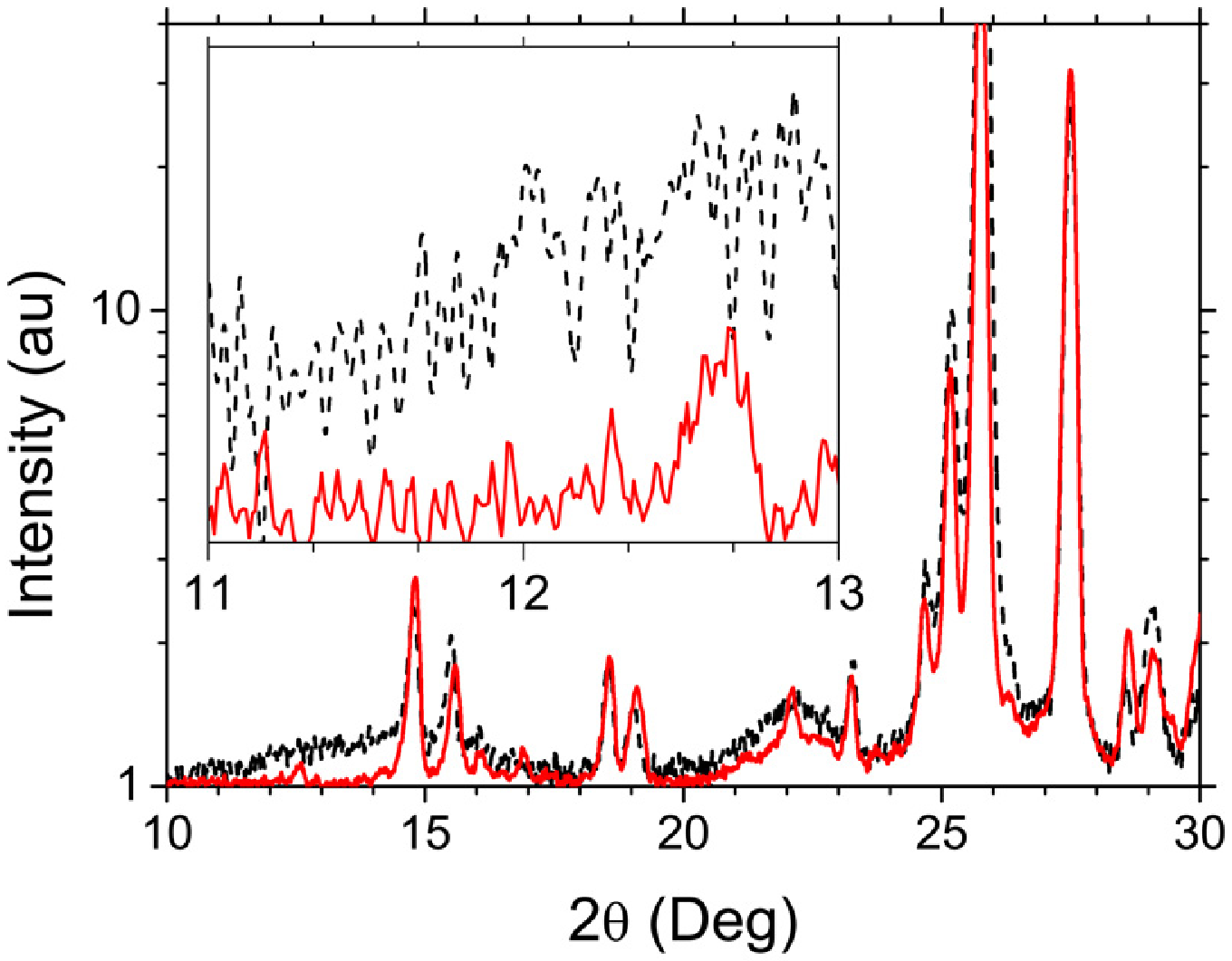}
\caption{X-ray diffraction patterns for the as-grown and annealed bilayer grown on Si(100)/SiO$_2$ substrate shown by dashed and solid lines, respectively. Inset: Angular range in which a small new Bragg peak appears at $12.66^{\circ}$ after annealing. See text for details. The intensity of the peaks is given in arbitrary units.}
\label{fig:X-ray_compar}
\end{figure}

To determine the crystal phase of the CT compound of (BEDT-TTF)-TCNQ we performed X-ray diffraction on the bilayers grown on glass and Si/SiO$_2$ substrates. After annealing we detected a new Bragg peak at $2\theta = 12.66^{\circ}$ which could be tentatively attributed to the $(10\bar{1})$ reflexion of the monoclinic (BEDT-TTF)-TCNQ phase (see Fig.~\ref{fig:X-ray_compar}). However, the small volume fraction of the black crystallites within the remaining bilayers of TCNQ and BEDT-TTF after annealing results in a correspondingly weak intensity of the X-ray diffraction pattern which is furthermore overlaid with the strong BEDT-TTF and TCNQ Bragg reflections. The volume fraction of the CT phase can be enhanced by co-evaporation, as we show next.
\subsection{Characterization of layers prepared by co-evaporation} \label{ssec:phase}
We prepared the layers by single-source evaporation from (BEDT-TTF)-\-TCNQ crystallites which we obtained from solution growth (sample type ``single-CT''), co-evaporation of TCNQ and BEDT-TTF (``co-A/D''), and co-evapora\-tion of TCNQ with (BEDT-TTF)-TCNQ (``co-A/CT''). The layers were grown in parallel on Si/SiO$_2$ and $\alpha$-Al$_2$O$_3$ (a-plane) substrates. We employed X-ray diffraction for phase analysis and made several observations.

The growth of ``co-A/D''-type samples is appreciably hindered by the fact that BEDT-TTF tends to dissociate for effusion cell temperatures above about 170~$^{\circ}$C. The higher effusion cell temperature is required because in the co-evaporation experiments the cell-to-substrate distance was by a factor of 16 larger than in the bilayer experiments. The BEDT-TTF dissociation became apparent from characteristic sulphur (0$\ell\ell$) Bragg reflexions, a color change of the source material from red-orange to dark brown and the occurrence of the specific smell of hydrogen sulphide H$_2$S in the vacuum pump exhaust. We note here that Knoll \textit{et al.} showed that BEDT-TTF fragments due to partial dissociation prevent an ordered growth of BEDT-TTF thin films \cite{knoll_apl}.
\begin{figure}[htb]
%{0.5\linewidth}
\includegraphics[width=1\textwidth]{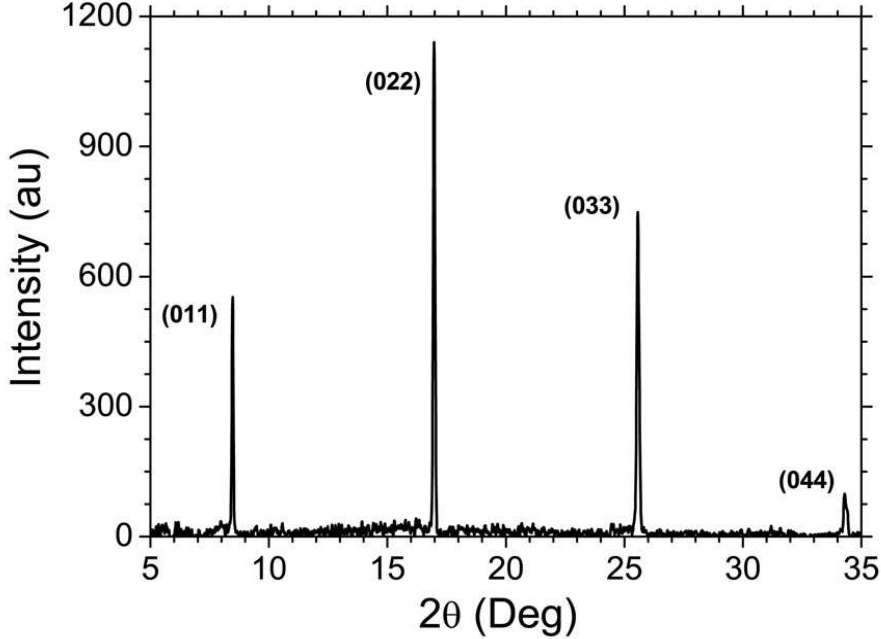}
\caption{X-ray diffraction pattern of ``co-A/CT''-type sample grown on $\alpha$-Al$_{2}$O$_{3}$ (a-plane) substrate kept at $100~^{\circ}$C. $(0\ell\ell)$ reflexions of BEDT-TTF can be identified up to fourth order. The intensity of the peaks is shown in arbitrary units.}
\label{fig:xrd-et}
\end{figure}

The growth of ``single-CT''-type samples on substrates held at room temperature resulted in the formation of thin, discontinuous films of the monoclinic CT phase of (BEDT-TTF)-TCNQ, albeit in badly crystallized form. For these experiments the effusion cell temperature was held at 136~$^{\circ}$C. The growth from the pre-reacted CT material was successfully applied to TTF-TCNQ (see for example Ref.~\cite{TCNQ_T}). In this case it was shown that the TTF-TCNQ CT phase formed easily, although it remained unclear whether the vapor contained donor-acceptor dimers or consisted of individual donor and acceptor molecules. For BEDT-TTF/TCNQ we consider it to be highly unlikely that the flux contains BEDT-TTF-TCNQ dimers because of the associated large mass. This assumption receives experimental support from the observation that pure, oriented and well-crystallized BEDT-TTF layers can be grown via the ``single-CT'' as well as ``co-A/CT'' route if the substrate is held at about 100~$^{\circ}$C. This is depicted in Fig.~\ref{fig:xrd-et} for a layer grown on $\alpha$-Al$_2$O$_3$ (a-plane) substrate. EDX measurements on a reference sample grown in parallel on Si/SiO$_2$ showed no trace of nitrogen. We are led to speculate that during evaporation the TCNQ flux component acts as a carrier gas for the BEDT-TTF molecules because TCNQ has an orders of magnitude larger vapor pressure at any given temperature. On the substrate surface the TCNQ then readily desorbs at elevated substrate temperature. Even if the substrates are held at room temperature we observed that TCNQ layers tend to desorb easily. With regard to the optimal route towards CT layer growth we conclude that the partial loss of TCNQ has to be compensated. This led us to perform the third type of co-evaporation experiment.

\begin{figure}[htb]
%{0.5\linewidth}
\includegraphics[width=1\textwidth]{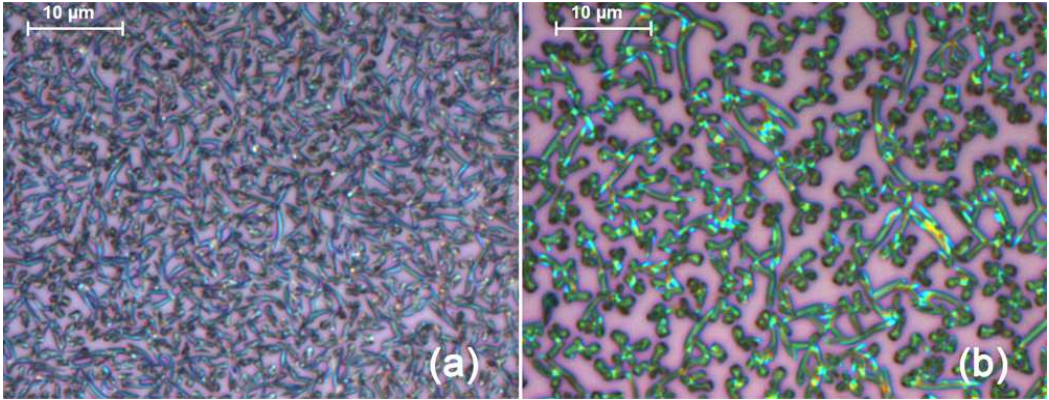}
\caption{Optical microscopy images of ``co-A/CT''-type samples grown on $\alpha$-Al$_{2}$O$_{3}$ (a-plane) (a) and on Si/SiO$_{2}$ (b) substrates taken in differential interference contrast. Average crystal size is about 4~$\mu$m and 6.5~$\mu$m for $\alpha$-Al$_{2}$O$_{3}$ (a-plane) and for Si/SiO$_{2}$ substrates, respectively. Images were taken at 500$\times$ magnification.}
\label{fig:coevapor}
\end{figure}

\begin{figure}[htb]
%{0.5\linewidth}
\includegraphics[width=1\textwidth]{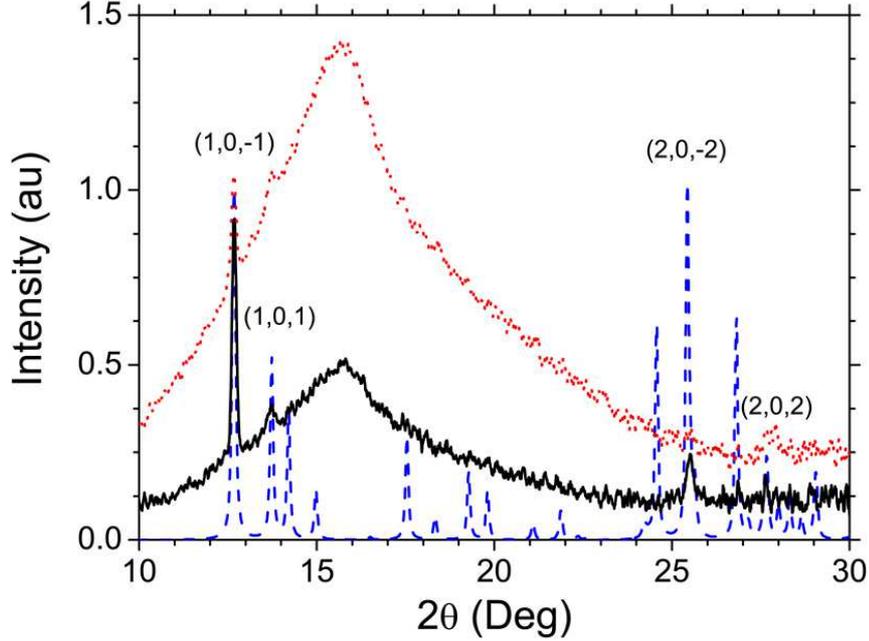}
\caption{X-ray diffraction pattern of (BEDT-TTF)-TCNQ samples of the ``co-A/CT''-type grown on Si/SiO$_{2}$ (dotted line) and on $\alpha$-Al$_{2}$O$_{3}$ (a-plane) (solid line). The intensity of the peaks is given in arbitrary units. Dashed line: Reference diffraction pattern simulated using Mercury 1.4.2 ``Crystal Structure Visualization'' \cite{Mercury} for the monoclinic phase of (BEDT-TTF)-TCNQ.}
\label{fig:X_ray}
\end{figure}

The growth of ``co-A/CT''-type samples resulted in the formation of better crystallized (BEDT-TTF)-TCNQ layers of the monoclinic variant. The layers were prepared at effusion cell temperatures of $130-138~^{\circ}$C for TCNQ and $136~^{\circ}$C for (BEDT-TTF)-TCNQ. Simultaneous thin film growth was done on $\alpha$-Al$_2$O$_3$ (a-plane) and Si/SiO$_2$ substrates, which were held at room temperature. Isolated small black crystals formed on the substrate surface (see Fig.~\ref{fig:coevapor}). We determined the sulphur-to-nitrogen ratio for a sample grown on Si/SiO$_2$ by EDX to 1.89$\pm$0.13, which is, within the error margin, in accordance with the expected ratio of 2.0 for the CT compound. The X-ray diffraction patterns for the samples appear in Fig.~\ref{fig:X_ray}. In the X-ray pattern of the sample grown on $\alpha$-Al$_{2}$O$_{3}$ (a-plane) substrate the peaks at $12.66^{\circ}$ and at $13.73^{\circ}$ correspond to the $(10\bar{1})$ and $(101)$ reflexions of the monoclinic phase of (BEDT-TTF)-TCNQ. The corresponding second order diffraction peaks occur at $25.53^{\circ}$ and $27.66^{\circ}$. For the Si/SiO$_{2}$  substrate only first order diffraction peaks can be resolved hinting towards bad crystallization of (BEDT-TTF)-TCNQ as compared to the other substrate material.

In the bilayer growth experiments there was no indication of CT phase formation at the donor/acceptor interface for the as-grown samples within the resolution of the electron microscope (about 4 nm at the given settings). Nevertheless, the co-evaporation route proved to be successful. We assume that this is due to the overall enlargement of the number of directly touching donor and acceptor molecules as compared to the bilayer structure. The released heat during the CT process may be sufficient to drive further reactions by enhancing the mobility of the molecules and helping to overcome the activation barrier for charge transfer.

We conclude this section by commenting on possible reasons why the monoclinic variant of the CT-phase was formed in our experiments. In the case of bilayer growth the CT-phase formation is close to equilibrium only at the interface region. For the co-evaporation the growth occurs with large supersaturation, that strongly suppresses the formation of CT-phase with the slow kinetics. However in both experiments the monoclinic phase of (BEDT-TTF)-TCNQ was detected. As we know from solution growth, the monoclinic variant of (BEDT-TTF)-TCNQ is formed in course of the fast growth, while if the process is slow enough $\beta$' phase of (BEDT-TTF)-TCNQ is more likely. Therefore the formation of $\beta$' phase is kinetically hindered as it also follows from physical vapor deposition growth, where the monoclinic variant is formed.
\section{Conclusion}
We studied the formation of a charge transfer phase in the (BEDT-TTF)-TCNQ donor-acceptor system prepared by physical vapor deposition. We used a donor/acceptor bilayer approach and performed comparative investigations of the layer interface for the as-grown and annealed bilayers. In this context focused ion beam cross sectioning delivered valuable information about the charge transfer phase formation initiated at the interface. Thus proven that a charge transfer phase formation can be initiated we also did evaporation experiments from the pre-reacted charge transfer phase, as well as co-evaporation experiments of the type donor/acceptor and acceptor/CT-phase. We found, first, that the pre-reacted CT phase can serve as source material for the preparation of oriented BEDT-TTF layers at elevated substrate temperature on $\alpha$-Al$_2$O$_3$ (a-plane). We consider this to be advantageous as compared to evaporation from pure BEDT-TTF source material and speculate that the TCNQ component acts as a carrier gas and thus allows for higher BEDT-TTF evaporation rates at moderate effusion cell temperature. Note, however, that further systematic experiments are needed to unequivocally show that TCNQ does indeed play such a carrier-gas role. Second, and more important for the present study, the co-evaporation of pre-reacted CT phase supplemented by TCNQ results in the direct formation of the monoclinic CT phase on substrates held at room temperature.

We believe that the bilayer approach used in our experiments can be a valuable method for a systematic study of charge transfer reactions in donor/acceptor systems prepared by the vapor phase approach. In particular, this approach can be extended to other donor/acceptor combinations. In order to improve the efficiency of the method finely modulated donor/acceptor multilayers can be applied which may serve as an intermediate of the bilayer and co-evaporation technique combining the advantages of both methods. Nevertheless, a crucial aspect in employing thin film structures is to optimize layer smoothness and layer orientation preference, ideally, towards epitaxial growth. In this regard, charge transfer phase formation studies, as performed here, are only the first step.

\section{Acknowledgment}
The authors acknowledge financial support by the DFG ($\it{Deutsche}$  $\it{For}$\-$\it{schungsgemeinschaft}$) through $\it{Sonderforschungsbereich/Transregio 49}$. They are also grateful to Roland Sachser for his support in doing the SEM experiments, Oleksandr Foyevtsov for his help and advice in the experiments and Dr. Ilia Solov'yov for fruitful discussions.

\end{document}